\documentclass{acm_proc_article-sp}
\usepackage[utf8]{inputenc}
\usepackage{xcolor}
\usepackage{graphicx}
\usepackage{amsmath}
\usepackage{url}
\usepackage{balance}
\graphicspath{ {img/} }

\title{Driving with Data: Modeling and Forecasting Vehicle Fleet Maintenance in Detroit}

\numberofauthors{7}
\author{
Josh Gardner \\
       \affaddr{University of Michigan}\\
       \email{jpgard@umich.edu}
\and
Danai Koutra \\
       \affaddr{University of Michigan}\\
       \email{dkoutra@umich.edu}
\and
Jawad Mroueh \\
      \affaddr{University of Michigan}\\
      \email{jmroueh@umich.edu} 
\and
Victor Pang \\
      \affaddr{University of Michigan}\\
      \email{vicpang@umich.edu}
\and
Arya Farahi \\
       \affaddr{University of Michigan}\\
       \email{aryaf@umich.edu}
\and
Sam Krassenstein  \\
       \affaddr{City of Detroit}\\
       \affaddr{Operations and Infrastructure Group}\\
       \email{samkrass@umich.edu}
\and
Jared Webb \\
       \affaddr{University of Michigan}\\
       \email{jaredaw@umich.edu} 
}

\date{June 2017}

\begin{document}

\maketitle

\begin{abstract}

The City of Detroit maintains an active fleet of over 2500 vehicles, spending an annual average of over \$5 million on new vehicle purchases and over \$7.7 million on maintaining this fleet. Understanding the existence of patterns and trends in this data could be useful to a variety of stakeholders, particularly as Detroit emerges from Chapter 9 bankruptcy, but the patterns in such data are often complex and multivariate and the city lacks dedicated resources for detailed analysis of this data. This work, a data collaboration between the Michigan Data Science Team\footnote{\url{http://midas.umich.edu/mdst}} and the City of Detroit's Operations and Infrastructure Group, seeks to address this unmet need by analyzing data from the City of Detroit's entire vehicle fleet from 2010-2017. We utilize tensor decomposition techniques to discover and visualize unique temporal patterns in vehicle maintenance; apply differential sequence mining to demonstrate the existence of common and statistically unique maintenance sequences by vehicle make and model; and, after showing these time-dependencies in the dataset, demonstrate an application of a predictive Long Short Term Memory (LSTM) neural network model to predict maintenance sequences. Our analysis shows both the complexities of municipal vehicle fleet data and useful techniques for mining and modeling such data.
\end{abstract}

\section{Introduction}

The City of Detroit, like many city governments, manages and maintains a large vehicle fleet consisting of over 2500 active vehicles. These vehicles support a diverse array of government functions, including service delivery, law enforcement, and grounds maintenance. In addition to being critical to Detroit's ability to effectively serve its citizens, the maintenance required to sustain this fleet is both complex and expensive: The city spent an annual average of \$7.7 million on maintenance and over \$5 million on new vehicle purchases between 2010 and 2017. As of 2015, the city had four shops, six fuel sites, and 70 technicians to maintain the fleet; this represented a significant fleet reduction after undergoing Chapter 9 bankruptcy filing in July 2013 \cite{White2015-zc}. Most cities lack the resources and expertise to dedicate to understanding and optimizing fleet operations, and even if those resources existed, analyzing the complex patterns of vehicle use and maintenance is a challenging task. Having a more nuanced understanding of the patterns in their fleet maintenance data would allow Detroit to make intelligent decisions for efficiency and cost-reduction at a critical time in the city's history. In particular, the city needs to balance concerns of cost and resource efficiency with maximizing vehicle uptime and lifetime, ensuring consistent service delivery, and reducing its carbon footprint.

\begin{figure}[t!]
    \centering
    \includegraphics[width = \columnwidth]{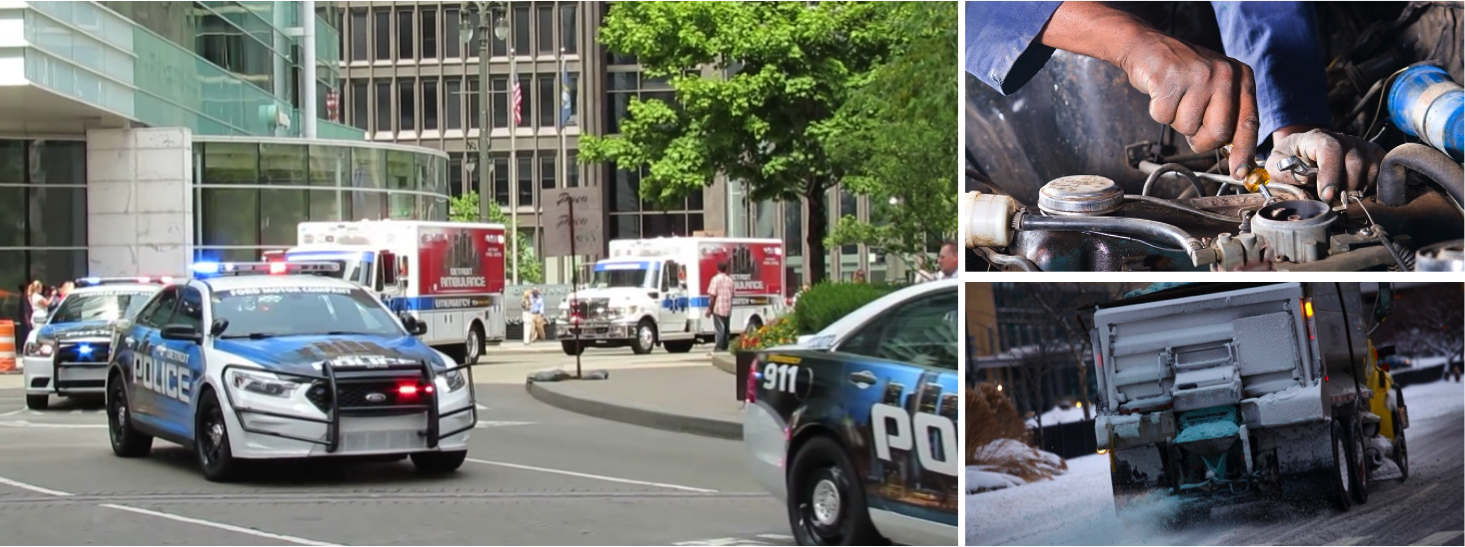}
    \vspace{-0.3cm}
    \caption{Vehicle fleet in the City of Detroit.}
    \label{vehicle-fig}
    \vspace{-0.4cm}
\end{figure}

This project, a data collaboration between the Michigan Data Science Team (MDST)---a student organization at the University of Michigan---, and the City of Detroit's Operations and Infrastructure Group, a municipal entity, is an initial foray into understanding and modeling municipal vehicle maintenance data. The analysis we present constitutes an initial step toward meeting the complex needs of the city (and citizens) of Detroit.

\begin{table*}[h] % force to top of page
\centering
\caption{Description of the \textit{vehicles} table.}
\label{vehicles-table}
{\small
\begin{tabular}{ p{3.75cm} p{7cm} p{3.75cm}}
\hline
{\bf Field} & {\bf Description} & {\bf Example} \\ \hline
Unit\# & Unique Vehicle Identifier & 026603 \\
Dept\# & Code of department vehicle is assigned to & 37 \\
Dept Desc & Description of department & POLICE \\
Make & Vehicle make & CHEVROLET \\
Model & Vehicle model & 2500 \\
Year & Model year of vehicle & 2002 \\
Last Meter & Odometer reading at last check (mi) & 52738 \\
Last Fuel Date & Most recent refuel at city refueling station & 2009-11-05 15:37:25 \\
Purchase Cost & Purchase cost, in US dollars & \$20,456 \\
Status Code & A = Active; S = Disposed & A \\
Status Desc & Description of status & Active Unit \\
LTD Maintenance Cost & Total maintenance cost to date, in US dollars  & \$5,951.04 \\
LTD Fuel Cost & Total fuel cost to date, in US dollars  & \$9,295.01 \\
LTD Fuel Gallons & Total fuel consumption to date & \$3,646.6 \\ \hline
\end{tabular}
}
\vspace{-0.4cm}
\end{table*}

The goals of this paper are two-fold: First, we aim to show that Detroit's fleet maintenance data contains discoverable structure, and to demonstrate methods for revealing this structure. Second, we seek to apply methods for modeling this structure to make predictions relevant to municipal decision-making and resource allocation, namely, forecasting vehicle maintenance. These predictions could reduce costs, fraud, and erroneous data; lead to better scheduling; and form the basis for future internal tools in the City of Detroit. In the analysis that follows, we pursue those aims by (a) exploring multidimensional patterns in vehicle maintenance using the parallel factors (PARAFAC) decomposition to reveal patterns in maintenance of automotive systems in different vehicle types over time using data \textit{tensors}, providing a visual approach to representing patterns in fleet maintenance both over time;  (b) applying a sequence mining technique to statistically identify frequent maintenance sequences by make/model, and (c) leveraging a modern neural network approach to predict vehicle maintenance in the City of Detroit's fleet. 

The structure of this paper is as follows: We first survey prior research on applied tensor analysis and on municipal vehicle fleets. In Section \ref{sec:data}, we describe the Detroit dataset in detail. Section \ref{sec:parafac} presents the results of the PARAFAC evaluation and in Section \ref{sec:mining} we conduct differential pattern mining to demonstrate the presence of statistically unique maintenance patterns by make/model and construct a predictive model utilizing a long short-term memory (LSTM) neural network to model these sequences. In Section 6, we give conclusions, challenges of data collaboration and analysis in real-world public-sector contexts. We conclude with suggestions for future work in Section 7.

\section{Related Work}
\label{sec:related}

Our analysis is based on tensor decompositions and related to other studies on municipal vehicle fleets. %We review these areas next.

\enlargethispage{\baselineskip}
\subsection{Tensor Analysis and Applications}

Tensor representations and various tensor decompositions have found wide applications in a variety of domains, including psychometrics \cite{Douglas_Carroll1970-jg} and brain imaging \cite{Mocks1988-al} (where many core techniques, such as the PARAFAC decomposition used here, were developed), the evolution of chatroom \cite{Acar2006-yl} and email \cite{Bader2008-rm} conversations over time, modeling web search \cite{Sun2005-dm}, epidemiology \cite{noauthor_undated-jb}, and anomaly detection \cite{Koutra2012-hx}. Tensor representation is useful in a variety of problem domains because it allows for multi-way analysis of data containing multidimensional patterns.

For a more detailed overview of tensor decompositions, their mechanics, and their applications, we refer the interested reader to \cite{Kolda2009-jh}. We describe the decompositions that are relevant to our data analysis in Section~\ref{section-tensor-rep}.

\subsection{Municipal Vehicle Fleets and Predictive  \\ Maintenance Models}

While predictive analytics, data science, and the application of such techniques to urban planning (sometimes called \textit{urban informatics}) have dramatically expanded in recent years, these techniques have seen only limited applications to one of the largest and most substantial assets managed by many governments---their vehicles---and published research on the topic is limited. Some state and local governments conduct, but rarely publish, fleet lifecycle reports and maintenance analyses \cite{Gransberg_undated-nb} . \cite{Lauria2014-ph} reports on fleet replacement management by state Departments of Transportation across all states. \cite{Osborne2012-ws} presents a case study of municipal fleet management in a mid-sized American city mostly focused on cost-reduction analysis.

Recent research on predictive maintenance has utilized on-board vehicle data to predict maintenance \cite{Prytz2014-na} the use of vehicle speed data to evaluate winter maintenance operations \cite{Lee2008-yz}. Other vehicle-related issues in urban areas have received significant research attention, including accident prediction \cite{Levine1995-cr, Levine1995-ea} and traffic flow prediction and optimization \cite{Vlahogianni2005-yn, Weizhong_Zheng2006-sf, Lv2015-li}. The authors are not aware of any prior research applying tensor decomposition or the other techniques used in the current work to municipal vehicle data.

\enlargethispage{\baselineskip}
\section{Dataset}
\label{sec:data}

In this section, describe the raw dataset obtained from the City of Detroit, and the transformation of the raw vehicle and maintenance data into the data tensor, to which we apply the tensor modeling techniques described in Section~\ref{sec:parafac}.

\subsection{Detroit Vehicles Dataset}

MDST partnered with the City of Detroit’s Operations and Infrastructure Group to obtain a comprehensive dataset from the City of Detroit. This dataset consists of two tables. 

The \textbf{vehicles table} consists of 6,725 records, one per vehicle, representing every known vehicle currently or previously owned by the City of Detroit. 2,566 of these vehicles are currently active in the fleet, but the oldest vehicle purchases date to 1944. The table includes information about each vehicle’s manufacture, purchase, and use. The table includes police cars, garbage trucks, freight trucks, ambulances, boats, motorcycles, mowers, and other vehicles. Table \ref{vehicles-table} gives a description of the fields, with a sample entry. %, is given in Table \ref{vehicles-table}.

The \textbf{maintenance table} consists of job-level records for all maintenance performed on any vehicles owned by the City of Detroit. This table includes 229,540 records representing individual jobs, which include everything from routine inspections, tire changes, and preventive maintenance to major collision repairs, glass work, upgrades, and engine replacements. The maintenance data is described in Table~\ref{maintenance-table}. 

\begin{table*}[] 
\centering
\caption{Description of the \textit{maintenance} table.}
\label{maintenance-table}
{\small
\begin{tabular}{p{4cm} p{6.5cm} p{4cm}}
\hline
{\bf Field} & {\bf Description} & {\bf Example} \\ \hline
Job ID & Unique identifier for job & 847956 \\
Year WO Completed & Year of completion & 2017 \\
Unit No & Vehicle identifier & 067602 \\
Work Order No & Unique identifier for work order & 635864 \\
WO Open Date & Work Order Open & 2017-01-17 \\
WO Completed Date & Work Order Completion & 2017-01-17 \\
Work Order Location & Location of work order & CODRF \\
Job Open Date & Job Open & 2017-01-17 \\
Job Reason & Job reason code & B \\
Job Reason Desc & Job reason description & BREAKDOWN / REPAIR \\
Job Open Date2 & Job Open 2 & 2017-01-17 \\
Job Completed Date & Job Completed & 2017-01-17 \\
Job Code & Job ID & 24-13-000 \\
Job Description & Detailed description of job & REPAIR Brakes \\
Labor Hours & Hours of labor completed on job & 6.35 \\
Actual Labor Cost & Cost of labor for job & \$348.16 \\
Commercial Cost & Cost of commercial (non-city) labor & \$0 \\
Part Cost & Cost of parts for job & \$57.55 \\
Primary Meter & Odometer reading at time of repair (mi) & 48250 \\
Job Status & Status code; DON = Done & DON \\
Job WAC & Job type code & 24 \\
WACDescription & Job type description & REPAIR \\
Job System & Code for vehicle system repaired by job & 13 \\
System Description & Description for vehicle system repaired by job & Brakes \\
Job Location & Location where job was completed & CODRF \\ \hline
\end{tabular}
}
\end{table*}

Together, these tables form a dataset representing detailed job-level information about maintenance on Detroit's entire vehicle fleet across 87 different departments, including police, airport, fire, solid waste, and grounds maintenance. There is no missing data, % in either table, 
but there are potential concerns about the accuracy of some data due to data-entry errors or human coding of job types and descriptions. Because the City of Detroit's fleet, and its data collection practices, have changed substantially over time, we limit our analysis to maintenance data from vehicles purchased in 2010 or later in order to utilize only the most reliable and  relevant vehicle and maintenance patterns. This represents 1,087 vehicles and over 25,000 individual maintenance records.

\subsection{Data Representation as a Tensor} \label{section-tensor-rep}

In this section, we describe the process of representing the Detroit vehicle maintenance dataset as a series of $vehicle \times system \times time$ data tensors, providing a brief introduction to tensors and describing both the process and motivation for this approach.

A tensor is a multidimensional or N-way array \cite{Kolda2009-jh}. Tensors provide a way of representing, analyzing, and modeling complex, multidimensional data. While tensors of arbitrary numbers of dimensions, or modes, can be evaluated using the techniques described here, the current analysis uses only 3-mode tensors which are, fortunately, straightforward to visualize and discuss.

In this analysis, we were interested in understanding how vehicle maintenance unfolded over time, and whether there were patterns and structure in how different types of vehicles were maintained. This task is of interest to our partners in Detroit in order to understand fleet maintenance, but also has the potential to inform future work on predicting vehicle maintenance, breakdowns, availability, and direct and indirect maintenance costs.

\begin{figure}[t!]
\vspace{-0.5cm}
    \centering
    \includegraphics[width = \columnwidth]{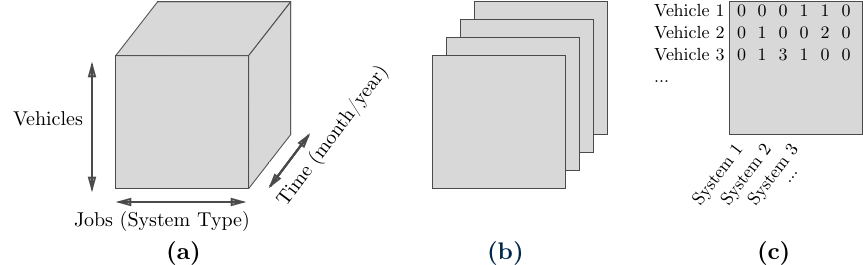}
    \vspace{-0.65cm}
    \caption{Depiction of (a) 3-mode data tensor; (b) the same tensor as a stacked series of frontal slices, or arrays; (c) an example single frontal slice of a vehicle data tensor used in this analysis (each entry corresponds to the count of a specific job type for a vehicle at a fixed time).}
    \label{tensor-fig}
    \vspace{-0.3cm}
\end{figure}

In order to represent the raw data -- which consisted of a \textit{vehicles} table and a \textit{maintenance} table -- as a data tensor, we needed to aggregate the data by vehicle, job type, and time. We assembled counts of maintenance jobs, by vehicle, system\footnote{\textit{System} describes the vehicle component repaired in a job, such as brakes, lighting system, or suspension; see ``System Description" in Table \ref{maintenance-table}.}, and month/year. This produced a 3-way tensor similar to the one shown in Figure \ref{tensor-fig}, where the vertical axis (the \textit{first mode}) represents each different vehicle, sorted by year and unit number; the horizontal axis (the \textit{second mode} represents each distinct vehicle system that occurred for at least one job in the dataset; and the depth (\textit{third mode}) represents time in months or years. The value at any given [vehicle, system, time] point in the tensor is the count of jobs for that particular vehicle, system, and month. An example of the data tensor we construct is shown in Figure \ref{tensor-fig}.

This 3-dimensional tensor representation allows us to model the relationships across these three dimensions in the data -- and in particular, to see how patterns evolve over time. This representation is critical to answering our initial question as to whether patterns exist in maintenance over time, which would lead directly to insights about maintenance trends in Detroit's vehicle fleet, inform approaches to modeling and prediction of vehicle maintenance, and potentially lead to changes in the city's fleet maintenance operations.

\section{Understanding Maintenance Patterns Over Time with PARAFAC}
\label{sec:parafac}

In this section, we describe a technique for extracting insights about structure and patterns in a data tensor known as the PARAFAC (PARAllel FACtors) decomposition, and describe insights gained from applying this technique to tensors of the vehicle fleet over absolute time, as well as over vehicles' lifetimes.

\subsection{The PARAFAC Decomposition}

Tensors can be thought of as higher-dimensional versions of the ``flat'' two-dimensional arrays common in data analysis tasks. As such, many techniques have been developed for manipulating and understanding tensors that are analogous to methods for two-dimensional data. The PARAFAC decomposition is an example of such a technique. PARAFAC decomposes a tensor into a sum of component rank-one tensors which best reconstruct the original tensor. Given a third-order tensor $\mathcal{X} \in \mathbb{R}^{I \times J \times K}$, PARAFAC decomposes the tensor as:
\begin{equation}
    \mathcal{X} \approx \sum_{r=1}^R \mathbf{a}_r \circ \mathbf{b}_r \circ \mathbf{c}_r
    \label{tensor-equation}
\end{equation}
where $R$ is a positive integer, $a_r \in \mathbb{R}^I$, $b_r \in \mathbb{R}^J$, $c_r \in \mathbb{R}^K$ for $r = 1, \cdots, R$  and ``$\circ$'' represents the vector outer product. Thus, PARAFAC represents each element of $\mathcal{X}$ as the vector outer product of the respective fibers of $\mathcal{X}$:
\begin{equation}
    x_{ijk} = \sum_{r=1}^R a_{ir}b_{jr}c_{kr} \small{\text{ for } i = 1,\cdots,I, j = 1,\cdots,J, k = 1,\cdots,K}.
\end{equation}
The PARAFAC decomposition can be written compactly as the combination of three loading matrices \textbf{A}, \textbf{B}, \textbf{C}:
\begin{equation}
    \mathcal{X} = [\mathbf{A}^{I\times R},  \mathbf{B}^{J\times R},  \mathbf{C}^{K\times R}],  
\end{equation}
in which the $r^{th}$ columns correspond to the vectors $\mathbf{a}_r$, $\mathbf{b}_r$ and $\mathbf{c}_r$, respectively. 
%
% \dk{we don't really need the following factors, right?} In three dimensions, the PARAFAC decomposition allows us to obtain a set of \textit{factor matrices} from the combination of the vectors from the rank-one components:
% \begin{equation}
% \begin{split}
%      \mathcal{X}_{(1)} \approx \textbf{A}(\textbf{C} \odot \textbf{B})^T \\
%     \mathcal{X}_{(2)} \approx \textbf{B}(\textbf{C} \odot \textbf{A})^T \\
%     \mathcal{X}_{(3)} \approx \textbf{C}(\textbf{B} \odot \textbf{A})^T
% \end{split}
%   \end{equation}
% %
% Where ``$\odot$'' represents the Khatri-Rao Product. 
For more information about the PARAFAC decomposition, we refer the reader to \cite{Kolda2009-jh}. The key aspect of the PARAFAC decomposition that makes it useful for understanding the Detroit vehicle-maintenance dataset is that it yields $R$ sets (components or factors) of \textbf{a}, \textbf{b}, and \textbf{c} vectors which best reconstruct the original data tensor. These $R$ factors can be thought of as containing the most ``important'' relationships between different fibers of the data tensor across all three dimensions (or modes).

\begin{figure}
    \centering
    \includegraphics[width = 8.5cm]{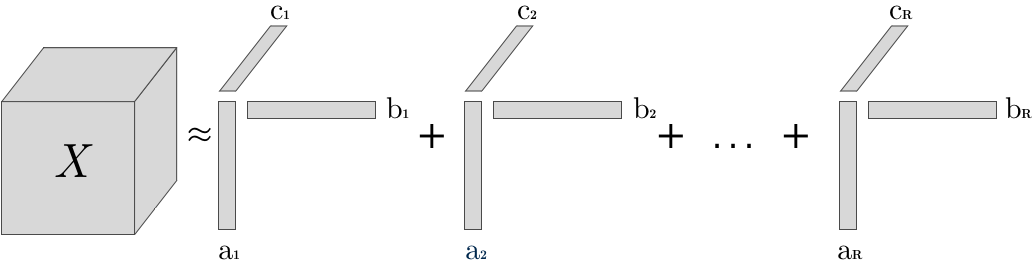}
    \caption{A visual depiction of the PARAFAC decomposition: It decomposes the $vehicle \times system \times time$ data tensor into vector products of the respective vehicle, system, and time dimensions. From this decomposition, we can obtain factor matrices, which reveal factors that are visualized to display relationships between these three variables in the Detroit vehicle-maintenance dataset. Figure based on \cite{Kolda2009-jh}.}
    \label{parafac-fig}
\end{figure}

\subsection{Insights From PARAFAC Application to Detroit Vehicle-Maintenance Dataset}

This analysis was intended to answer the question of whether vehicle-system-time relationships exist in the Detroit dataset. To this end, we generate so-called ``three-way'' plots of the three factor matrices from the PARAFAC decomposition \cite{Koutra2012-hx}, using the MATLAB tensor toolkit provided by \cite{Bader2007-si, Bader_undated-ue} for PARAFAC. Each plot visualizes the vectors $\mathbf{a}_r$, $\mathbf{b}_r$ and $\mathbf{c}_r$, which correspond to the $r^{th}$ factor and represent the different modes participating in the factor (i.e., vehicle, system description and time, respectively). We explored two different representations of time for the third mode of the data tensors: one which used \textit{absolute time} (month and year, January 2010 - present, measured by the start date of a maintenance job) and another using \textit{vehicle lifetime} (year, starting with year 0 as the vehicle's purchase year). The absolute time analysis allows us to model seasonality and other real-time trends in fleet maintenance, and could be more useful in forecasting future maintenance, while the vehicle lifetime analysis allows us to measure trends and changes in vehicles' maintenance over the course of their lifetime in the Detroit fleet, and could be useful for vehicle make/model reliability analyses. We examine $R = 25$ components for each analysis.

\begin{figure}
    \centering
    \includegraphics[width=3.4in]{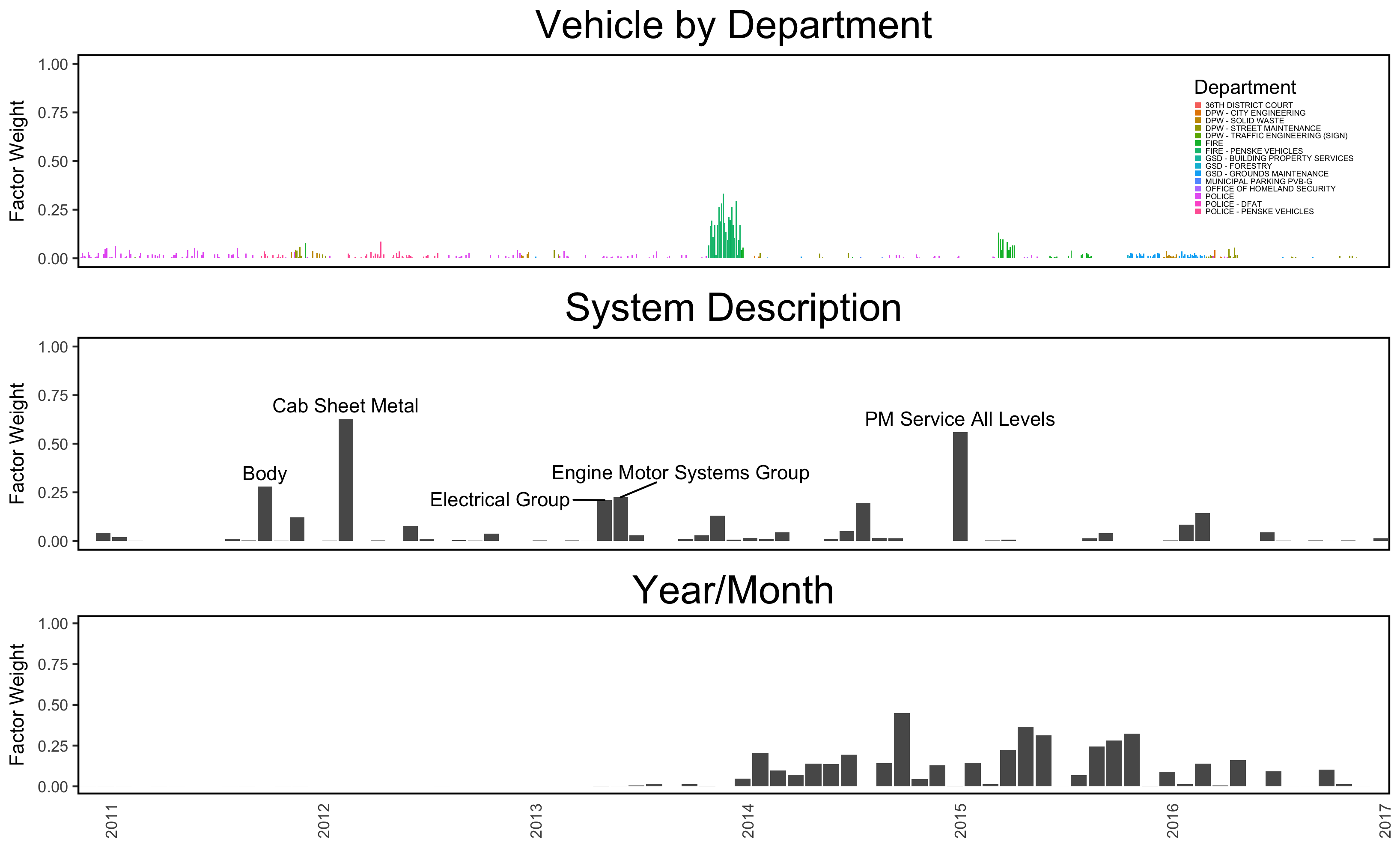}
    \caption{PARAFAC 3-way plot of absolute-time analysis. High factor weights in the top panel are for 2014 Terrastar Horton vehicles, an ambulance. The bottom two panels show systems (Body, Cab/Sheet Metal, Engine and Motor, and Preventive Maintenance Service) and time frames where this maintenance most often occurs.}
    \label{factor-fig-1}
    \includegraphics[width=3.4in]{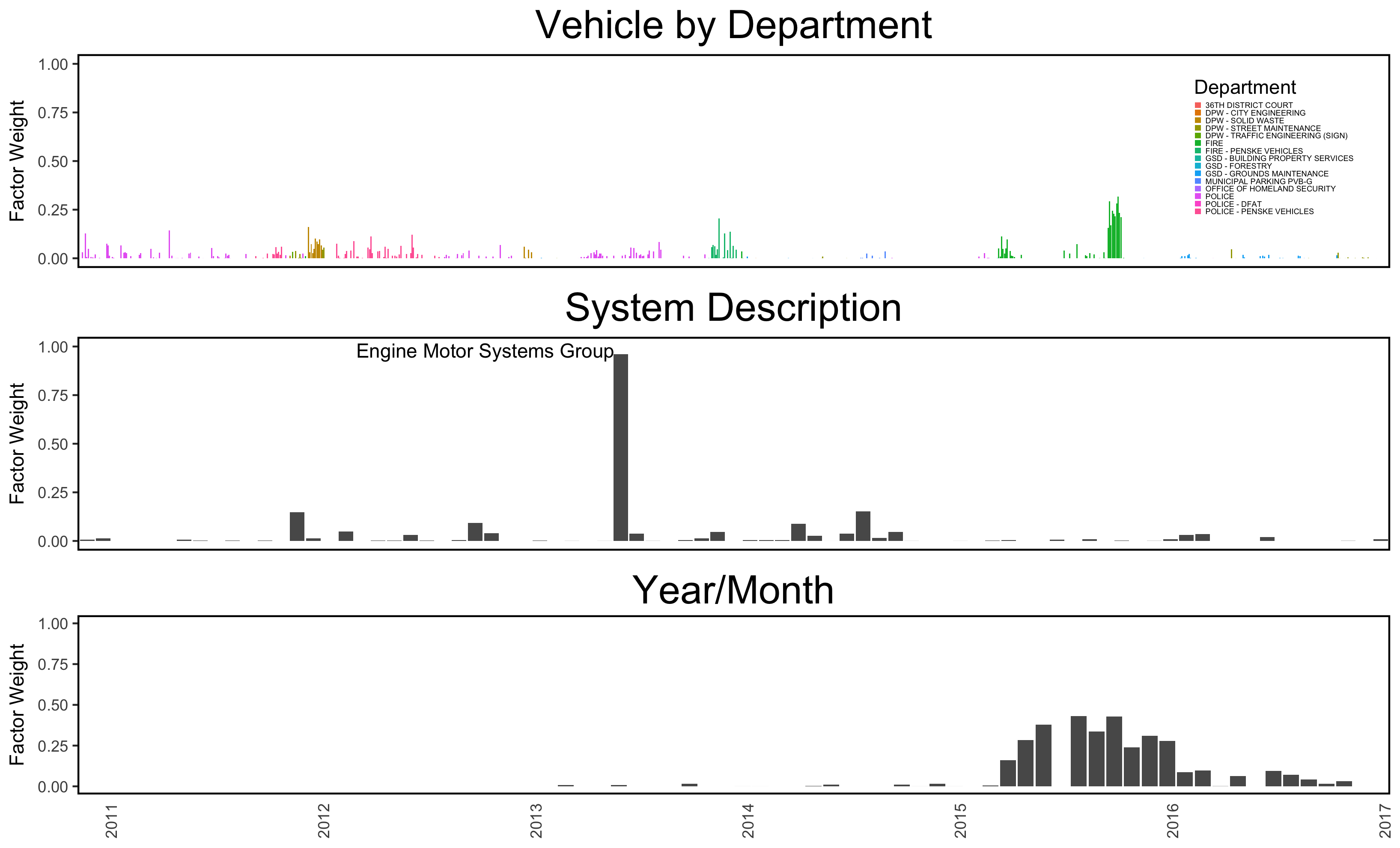}
    \caption{PARAFAC 3-way plot of absolute-time analysis. This  plot demonstrates strong and specific maintenance patterns for the 2015 Smeal SST Pumper fire truck. It shows extensive and specific repair to the engine systems with little other maintenance, from late 2015 through 2016.}
    \label{factor-fig-2}
    \includegraphics[width=3.4in]{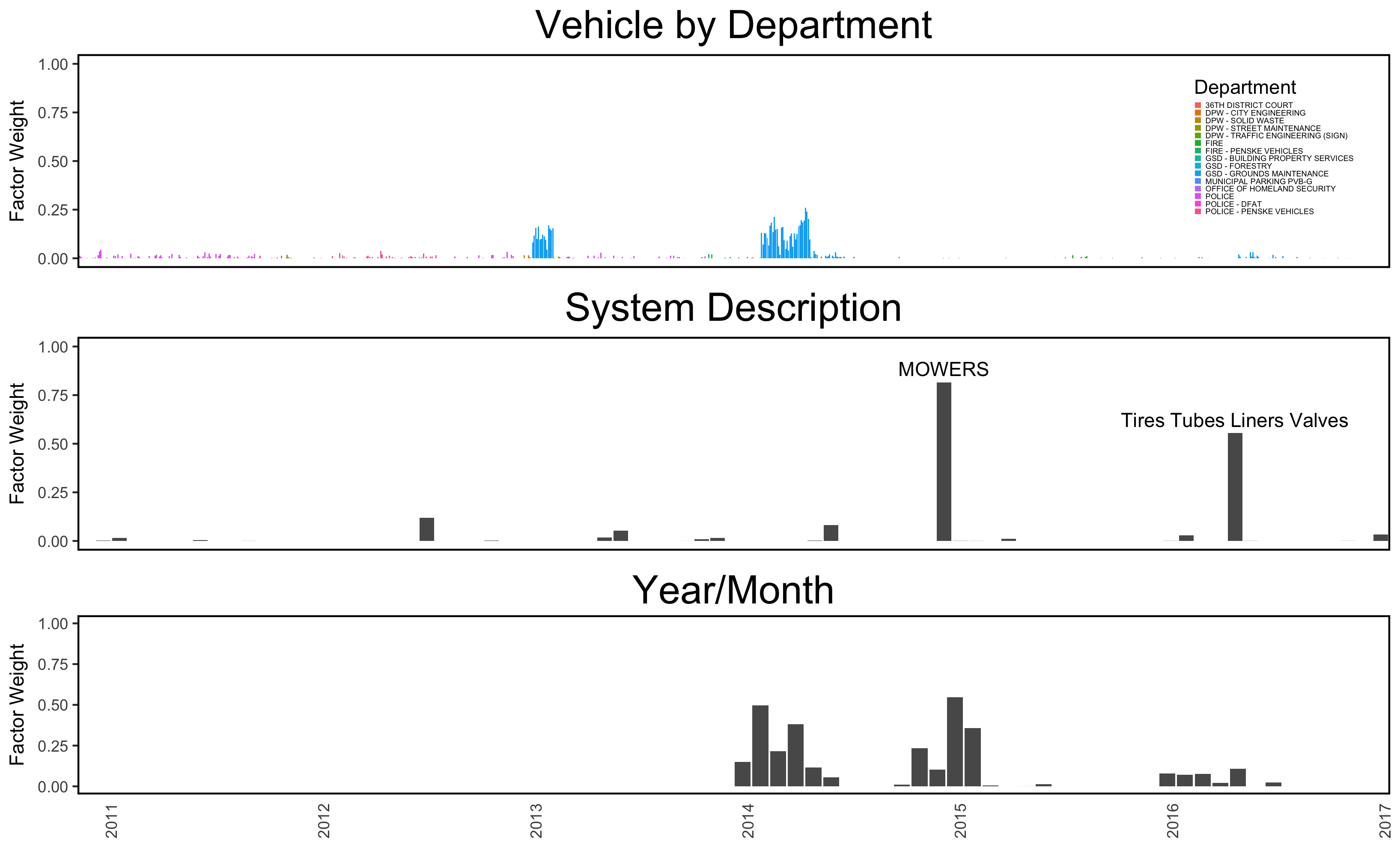}
    \caption{PARAFAC 3-way plot of absolute-time analysis. This plot shows factors related to the 2014 Hustler X-One, a Grounds Maintenance Department riding mower. There are strong system and time patterns to X-One maintenance, with repairs to mower blades and tires/tubes/liners/valves during seasons of high usage.}
    \label{factor-fig-3}
\end{figure}

Examples of the results from the absolute time analysis are shown in Figures \ref{factor-fig-1} - \ref{factor-fig-3}. These results demonstrate clear patterns across vehicles, systems under repair, and time, underscoring the importance of taking a multivariate approach. For example, fire trucks and ambulances (the Terrastar Horton in Figure \ref{factor-fig-1} and Smeal SST Pumper in Figure \ref{factor-fig-2}, respectively) both show strong evidence of patterns in their maintenance, but with very different groups of systems and across different time bands. The riding mower owned by the GSD - Grounds Maintenance Department shown in Figure  \ref{factor-fig-3}, however, displays an entirely different maintenance pattern, with a focus on only two systems (mowing blades and tires/tubes/liners) and strong seasonality which reflects the seasonal use of mowers in a northern city such as Detroit.

\begin{figure}
    \centering
    \includegraphics[width=3.4in]{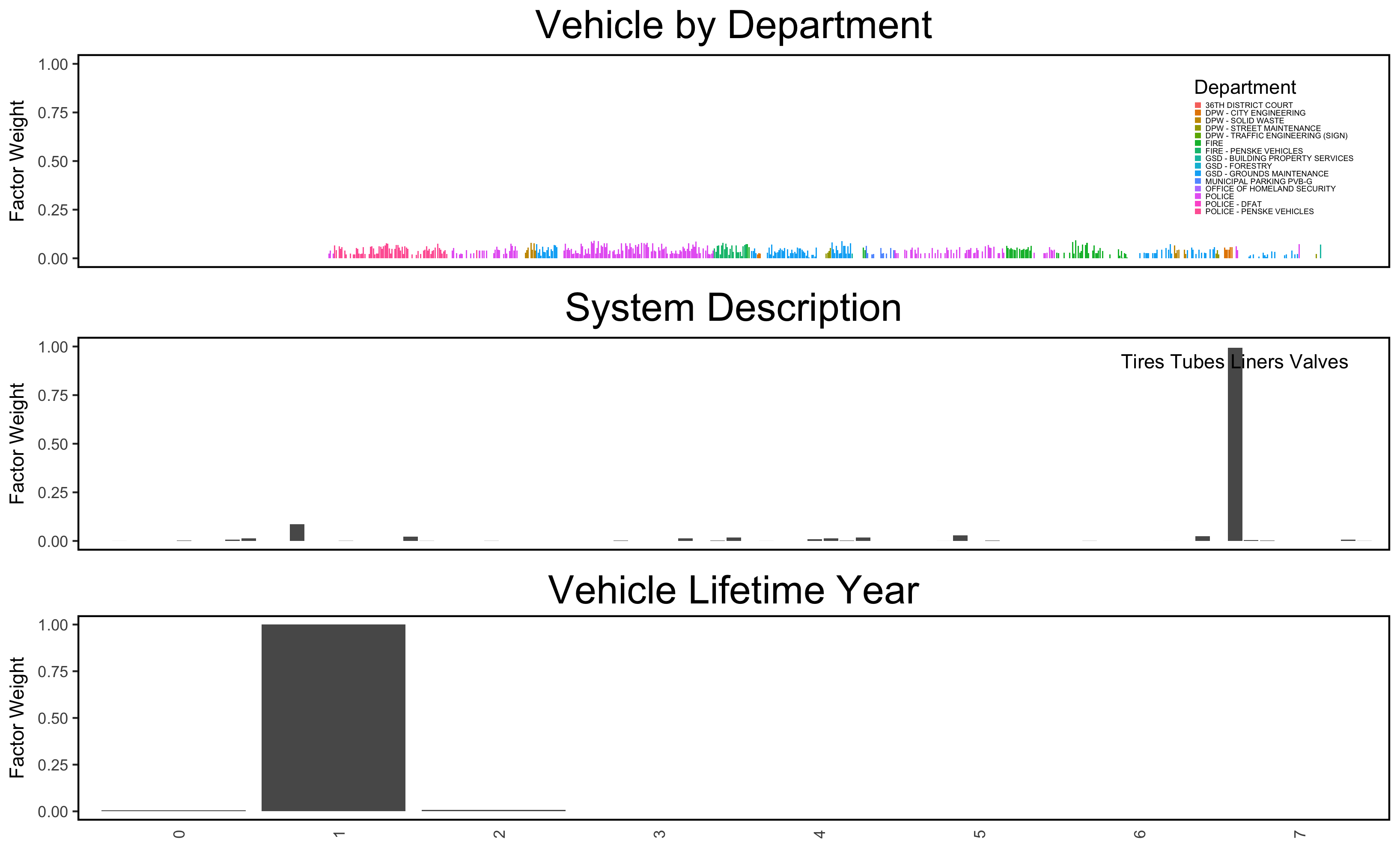}
    \caption{PARAFAC 3-way plot of vehicle lifetime analysis revealing a simple pattern common to almost \textit{all} vehicles, as demonstrated by the consistent loading across the vehicle factor (top panel): tires/tubes/valves/liners replacement during the second year of lifetime, with few repairs to this system either before or after.}
    \label{factor-fig-4}
    \includegraphics[width=3.4in]{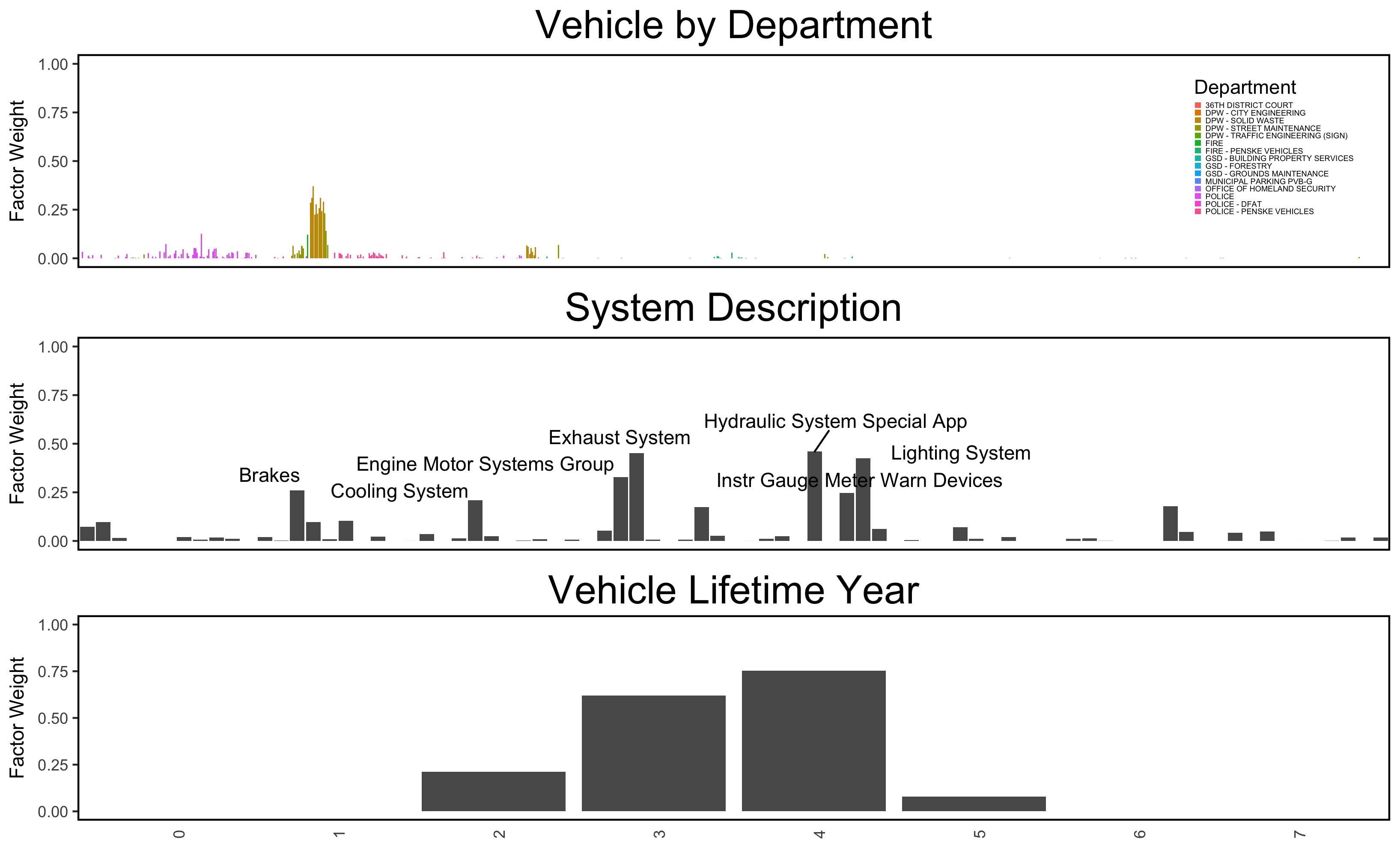}
    \caption{PARAFAC 3-way plot of vehicle lifetime analysis showing the 2012 Freightliner M2112V, a Department of Solid Waste garbage truck. This plot reveals a strong pattern of increased maintenance in years 2-4 after purchase, focusing on a variety of technical systems: hydraulics, lighting, gauges and warning devices, and cooling systems.}
    \label{factor-fig-5}
    \includegraphics[width=3.4in]{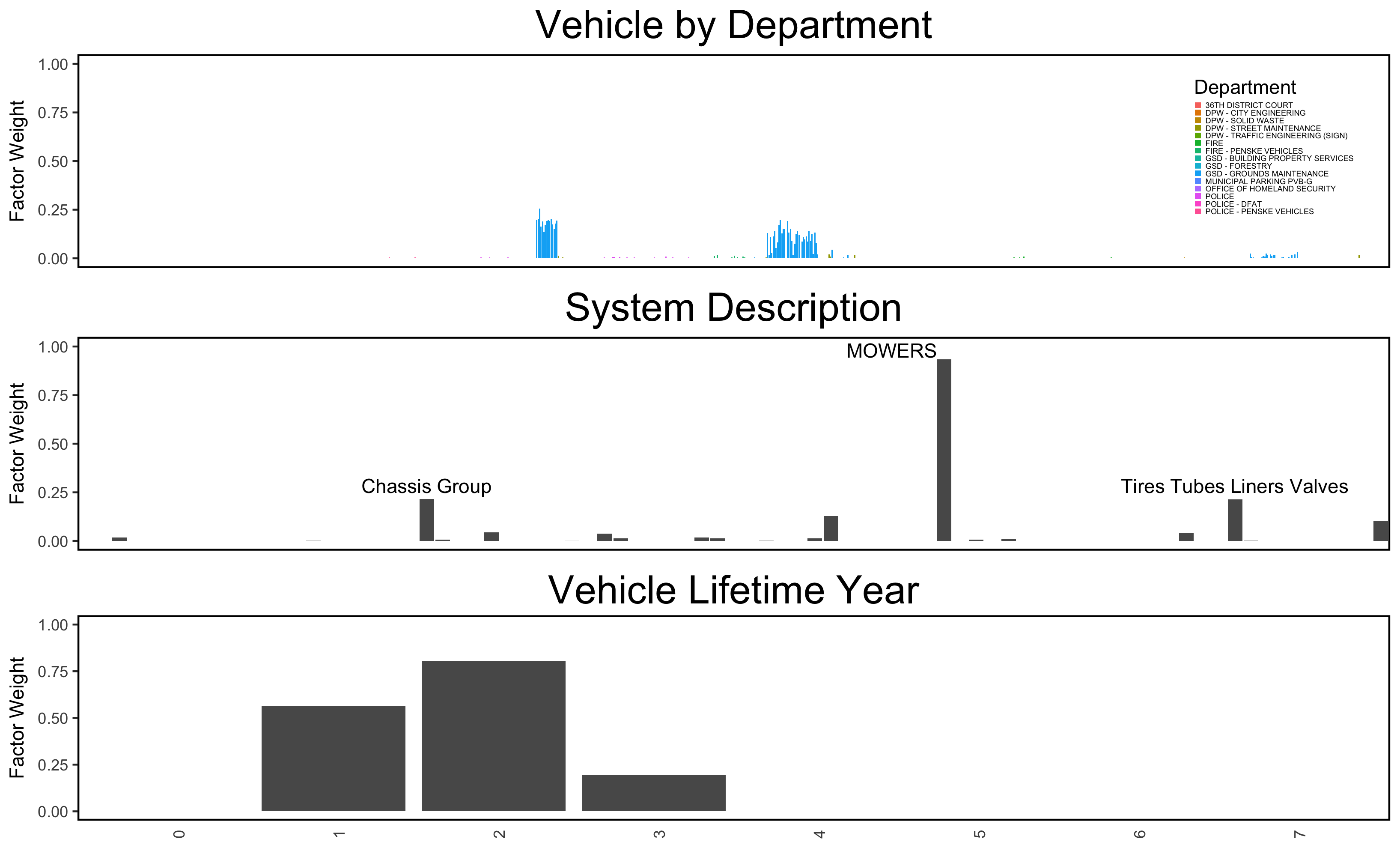}
    \caption{PARAFAC 3-way plot of vehicle lifetime analysis. This plot demonstrates maintenance patterns for the 2013 Hustler Z 60 2013 (a riding mower), which have mowing blades serviced frequently in the second and third years of their lifetime.}
    \label{factor-fig-6}
\end{figure}

Examples of the results from the PARAFAC vehicle lifetime analysis are shown in Figures \ref{factor-fig-4} - \ref{factor-fig-6}. This analysis demonstrates a different set of patterns -- this time, across the lifetime of vehicles, beginning when they are purchased.

\enlargethispage{\baselineskip}
The PARAFAC analysis via three-way factor plots demonstrates the variety of insights that can be gained from using tensor decomposition to understand complex multidimensional data. The analysis shown above reveals common trends across the entire Detroit vehicle fleet, as well as unique trends specific to certain vehicles, systems, and times. Additionally, the use of two different measures of time---month/year, and vehicle lifetime---allows us to demonstrate two different modes of time-bound pattern in the data. Such an approach demonstrates that there are unique patterns in the Detroit vehicle-maintenance dataset by vehicle, system, and time, suggesting that analysis and modeling approaches which can capture these patterns are likely to be effective. 

% \begin{figure}
%     \includegraphics[width = \columnwidth]{smeal_sst_pumper_crop}
%     \caption{2015 Smeal SST Pumper in Detroit.}
% \end{figure}

\section{Mining Frequent Maintenance \\ Patterns}
\label{sec:mining}

In Section \ref{sec:parafac}, our results demonstrate discoverable structure in the Detroit vehicle-maintenance data, particularly by vehicle make/model. In this section, we expand on these results, statistically verifying the existence of unique patterns in the \textit{sequences} of systems repaired by vehicle make and model and applying a sequence modeling approach to build a predictive maintenance model.

\subsection{Sequence Mining By Vehicle Make/Model}
\sloppy
Sequential pattern mining is a constellation of techniques used to identify and evaluate sequences of events \cite{Agrawal1995-qd}. \textit{Differential} sequence mining compares differences in sequences between two groups, statistically identifying different pathways unique to each group. We apply a methodology adapted from \cite{Kinnebrew2013-ny} to a subset of vehicles identified as having potentially unique maintenance patterns in the tensor decomposition analysis above, both in order to statistically verify these unique patterns, and to determine whether we can ignore time (and simply focus on order) in modeling maintenance sequences. Specifically, the method consists of three steps: 
%
% We apply the methodology from \cite{Kinnebrew2013-ny} in the following steps: 
\vspace{-0.3cm}
\begin{itemize}
\item \textbf{Step 1:} Find the top $n$ most frequent sequences (restricting these to sequences of length 3 or longer) for a given make/model and normalize by the total number of maintenance sequences of the same length for that make/model (this generates the \textit{left support} and the \textit{left normalized support}), using a general algorithm to find the most frequent sequences \cite{Wang2007-qp}.

\item \textbf{Step 2:} Calculate the same ratio for all other make/models as a separate group (the \textit{right support} and \textit{right normalized support}).

\item \textbf{Step 3:} Compare these two normalized frequencies by (a) calculating the left:right ratio, the \textit{i-ratio} \cite{Kinnebrew2013-ny}, and (b) by conducting a test of the difference between two population proportions to test the null hypothesis $H_0: p_1 = p_2$ against $H_a: p_1 \neq p_2$ where $p_1$ and $p_2$ are the left and right normalized supports, respectively.\footnote{Note that in the original implementation, a $t$-test was used; here, we use a difference-in-proportions $z$-test because our analysis tests whether the \textit{normalized} support differ, not the raw counts \cite{Kinnebrew2013-ny}.}
\end{itemize}

% The difference-in-proportions test uses the test statistic:

% \begin{equation}
%     z = \frac{p_1 - p_2}{\sqrt{p(1-p)(\frac{1}{n_1}+\frac{1}{n_2})}}
% \end{equation}

% To test the null hypothesis $H0: p_1 = p_2$ against $H_a: p_1 \neq p_2$, where p is the pooled sample proportion, and $n_1$ and $n_2$ are the size of the two samples under comparison.

The result of this analysis is shown in Table \ref{sequence-table}. These results demonstrate strong and statistically significant distinctions in maintenance sequences by vehicle type, suggesting that common maintenance sequences for these vehicles are unique to their make/model and statistically uncommon across the rest of the fleet. All sequences evaluated for the Dodge Charger, Ford Crown Victoria, and Smeal SST Pumper exceed any reasonable significance threshold, with only the Hustler X-One demonstrating less significant results for three of seven sequences tested (due to similarity with other models of Hustler mowers)\footnote{Note that we report $p$-values unadjusted for multiple comparisons, as this test is a heuristic to search for differences in patterns  as in \cite{Kinnebrew2013-ny} and not a strict statistical test; however, even conservative $p$-value adjustments would still yield highly significant results in most cases.}. Furthermore, because the sequential pattern mining approach simply looks at the \textit{order} of the maintenance jobs, but not their actual timing, these results demonstrate that there is strong correlation between maintenance sequence and make/model, even when actual timing is ignored, and that sequential models---even those which ignore the time between maintenance events---may be effective in modeling vehicle maintenance. 
This analysis informs the approach adopted in the following section.

\begin{table*}[ht!]
\centering

\caption{Results of differential sequence mining for selected make/models from Section \ref{sec:parafac}. The table shows the top 5 most frequent maintenance sequences of length $\geq$ 3 for each make/model (`Left'), which are compared to the frequency of that maintenance sequence across the rest of the fleet (`Right'). Large i-ratios and $z$-statistics indicate sequences unique to a given make/model. Abbreviations: TTLV = Tires, Tubes, Liners \& Valves; PM = PM Service All Levels; EX = Exhaust; MOW = MOW; ENG/MS = Engine / Motor Systems Group; PUMP = Pump - Product Transfer; CSM = Cab \& Sheet Metal.}

\label{sequence-table}
{\small 
\begin{tabular}{p{1cm} p{4.3cm} p{1.3cm} p{1.6cm} p{1.3cm} p{1.8cm} p{1.1cm} p{0.8cm} p{1.2cm}}

\hline
{\bf Vehicle} & {\bf Sequence} & {\bf Left Support} & {\bf Left Norm Support} & {\bf Right Support} & {\bf Right Norm Support} & {\bf i-Ratio} & {\bf z} & {\bf P(z)} \\ \hline
Dodge & (PM, TTLV, PM) & 187 & 0.0377 & 126 & 0.0067 & 5.6 & -10.4 & < 0.0001 \\
Charger & (PM, PM, TTLV) & 186 & 0.0375 & 81 & 0.0043 & 8.67 & -9.9 & < 0.0001 \\
 & (PM, PM, PM) & 185 & 0.0373 & 97 & 0.0052 & 7.2 & -10.3 & < 0.0001 \\
 & (TTLV, PM, PM) & 185 & 0.0373 & 82 & 0.0044 & 8.51 & -10.1 & < 0.0001 \\
 & (PM, TTLV, TTLV) & 183 & 0.0369 & 158 & 0.0085 & 4.37 & -11.3 & < 0.0001 \\
 & (TTLV, TTLV, PM) & 183 & 0.0369 & 168 & 0.009 & 4.11 & -11.4 & < 0.0001 \\
 & (TTLV, PM, TTLV) & 182 & 0.0367 & 180 & 0.0096 & 3.82 & -11.7 & < 0.0001 \\
 & (PM, TTLV, PM, TTLV) & 180 & 0.0378 & 40 & 0.0022 & 17.03 & -9.0 & < 0.0001 \\ \hline
Ford & (TTLV, PM, TTLV) & 101 & 0.0247 & 365 & 0.0187 & 1.32 & -18.4 & < 0.0001 \\
Crown & (PM, TTLV, TTLV) & 99 & 0.0242 & 333 & 0.017 & 1.42 & -18.6 & < 0.0001 \\
 Victoria & (PM, PM, TTLV) & 99 & 0.0242 & 130 & 0.0066 & 3.64 & -19.3 & < 0.0001 \\
 & (TTLV, TTLV, TTLV) & 99 & 0.0242 & 285 & 0.0146 & 1.66 & -18.6 & < 0.0001 \\
 & (PM, TTLV, PM) & 97 & 0.0237 & 248 & 0.0127 & 1.87 & -18.9 & < 0.0001 \\
 & (TTLV, TTLV, PM) & 97 & 0.0237 & 295 & 0.0151 & 1.57 & -18.8 & < 0.0001 \\ \hline
Hustler & (MOW, MOW, TTLV) & 49 & 0.0486 & 37 & 0.0016 & 29.72 & -1.6 & 0.1128 \\
X-One & (MOW, MOW, MOW) & 48 & 0.0476 & 70 & 0.0031 & 15.39 & -2.4 & 0.0149 \\
 & (MOW, TTLV, MOW) & 48 & 0.0476 & 39 & 0.0017 & 27.62 & -2.1 & 0.0331 \\
 & (MOW, TTLV, TTLV) & 48 & 0.0476 & 28 & 0.0012 & 38.47 & -2.0 & 0.0442 \\
 & (TTLV, MOW, MOW) & 47 & 0.0466 & 34 & 0.0015 & 31.02 & -2.6 & 0.0088 \\
 & (MOW, MOW, MOW, MOW) & 47 & 0.049 & 36 & 0.0017 & 29.7 & -1.3 & 0.1841 \\
 & (MOW, MOW, TTLV, TTLV) & 47 & 0.049 & 9 & 0.0004 & 118.81 & -0.9 & 0.3905 \\ \hline
Smeal & (EX, EX, EX) & 12 & 0.0198 & 11 & 0.0005 & 41.41 & -24.0 & < 0.0001 \\
SST & (EX, 'PUMP', EX) & 11 & 0.0181 & 0 & 0.0000 & 10000.0 & -36.0 & < 0.0001 \\
Pumper & (EX, EX, EX, EX) & 11 & 0.0185 & 3 & 0.0001 & 136.93 & -30.7 & < 0.0001 \\
 & (CSM, EX, EX) & 11 & 0.0181 & 2 & 0.0001 & 208.78 & -33.2 & < 0.0001 \\
 & (CSM, EX, EX, EX) & 11 & 0.0185 & 0 & 0.0000 & 10000.0 & -34.5 & < 0.0001 \\
 & (ENG/MS, EX, EX) & 11 & 0.0181 & 7 & 0.0003 & 59.65 & -28.4 & < 0.0001 \\
 & (ENG/MS, EX, EX, EX) & 11 & 0.0185 & 3 & 0.0001 & 136.93 & -30.7 & < 0.0001 \\ \hline

\end{tabular}
}
\vspace{-0.3cm}
\end{table*}

\subsection{Predicting Maintenance Sequences}

In this section, we build off of the findings of our prior analysis to construct a predictive model to predict the next maintenance job, given a vehicles' previous jobs, by learning from the maintenance histories of similar make/models.

Having demonstrated strong sequential patterns in maintenance for vehicle make and models, we developed an exploratory model to predict vehicle maintenance -- one of the potential applications of our data collaboration identified by Detroit's Operations and Infrastructure Group. From the raw data, we assemble a dataset consisting of the complete sequence of system repair jobs for each vehicle. Each vehicle's sequence is considered a separate observation. We train a probabilistic model which assigns probabilities to various repair sequences using a Long Short-Term Memory (LSTM) neural network \cite{Hochreiter1997-jm}. We specifically implement the architecture used in  \cite{Zaremba2014-vx} for predicting words in sentences because of its ability to model complex sequences while avoiding overfitting.

An LSTM model reads over a sequence, one item at a time, and computes probabilities of the possible values for the next item. In theory, an LSTM is capable of learning arbitrarily long-distance dependencies across a sequence; in our implementation, the LSTM considers a step size of 20 items. This means that the model considers up to 20 previous items in the sequence (if they exist) when predicting the next job. This model uses a dense representation of the input features, which allows it to learn about relationships between repairs to different systems. A feature that makes this model particularly well-suited to our problem is that it utilizes a technique called \textit{dropout}~\cite{Zaremba2014-vx} to regularize the model and avoid overfitting, which allows the network to model the vehicle data accurately without learning spurious or irrelevant patterns in the relatively small training dataset.

To assemble training, validation, and testing datasets for the model, we use all data from three make/models all often used as police cars, with similar maintenance patterns: Dodge Charger, Ford Crown Victoria, and Chevrolet Impala (see Table \ref{sequence-table} for an illustration of the similarities between frequent repair sequences for the Charger and Crown Victoria). Ideally, a model would be fit on only a single make/model; however, due to the relatively small size of the dataset (in total, this consists of only 329 vehicles), it was necessary to combine multiple make/models. We assemble the repair sequences for each vehicle and train on a random subset of 50\% of these vehicles, using 25\% for model validation and 25\% for testing.

% Again, following \cite{Zaremba2014-vx}, 
We assess the performance of our model using average per-item perplexity, a common evaluation metric for sequence models which evaluates the probability assigned to entire test sequences (an effective model would assign a high probability to unseen data): 
\begin{equation} %\label{eq:perplexity}
    e^{-\frac{1}{N} \sum_{i=1}^N\ln(p_{{\rm target}_i})} = e^{\rm loss}
    \label{eq:perplexity}
\end{equation}
The performance of our model, which achieves an average test perplexity score of 15.7, demonstrates that even this relatively simple, computationally lightweight model with a small dataset is able to achieve a reasonable performance on testing data. While perplexity benchmarks vary considerably by task, we can compare this with the perplexity of a `random' model which assigns an probability proportional to the frequency of item in the list to any given sequence of the 50 different system types observed in the training data. According to Equation \eqref{eq:perplexity} such a model would achieve a test perplexity of $260 \pm 40$, substantially larger than the LSTM model. We can also compare these results to the original application of our model, which achieved perplexity of 23.7 on the Penn Treebank dataset, and the state-of-the-art performance benchmarks on the Google Billion Words dataset \cite{Chelba2013-no, Shazeer2017-bi, Kuchaiev2017-ip} (perplexity of 43.8, 28.0, and 24.29, respectively). We note that our model's low perplexity score cannot be directly compared to model performance on other corpora, however it reflects the relatively high degree of predictability (and the relatively low number of candidate items in the sequence -- 81 unique systems in the entire vehicles dataset compared to many thousands in text corpora).

\enlargethispage{-2\baselineskip}
\section{Conclusions and Challenges}
\label{sec:conclusions}

In this analysis, we describe the initial results of a data collaboration between MDST and the City of Detroit's Operations and Infrastructure Group. This work demonstrates that there is significant, but highly complex, structure in the City of Detroit's vehicle-maintenance data. The complexity arises from  inter-relationships between vehicle type, system repair type, and time (both absolute time and vehicle life time). We employ PARAFAC tensor decomposition to uncover and visualize these relationships. Sequential pattern-mining adapted in this work verifies the time-vehicle relationship, and we find a statistically significant sequential patterns in maintenance. We note that a predictive model can accurately capture this sequence structure to make effective predictions using the available (modest-in-size) data. %, despite its modest size.

This collaboration demonstrates a small sample of the insights that can be gained from detailed multivariate analysis of municipal data, and illustrates several of the challenges of working with such data. Many aspects of the data---its observational nature; overlapping or difficult-to-decipher descriptions; error and incompleteness which are likely systematic and nonrandom\footnote{For example, technicians subjectively choose between several job codes: i.e., ``Adjust brakes'' vs. ''Repair brakes'' vs. ``Overhaul brakes''; many older vehicles and jobs are believed to be missing from this data.}---underscore the challenges of working with real-world municipal data often generated as ``data exhaust'' and not with the express aim of providing insights or accurate measurements. Additionally, the disconnect between our analytical team and the users generating the data (vehicle drivers, technicians, and clerical staff) highlights %the challenges of understanding the data context. 
how challenging it can be to understand data context.

The tools used and generated in this analysis are  open-source, including the MATLAB code used to generate the PARAFAC decompositions \cite{Bader2007-si, Bader_undated-ue} and the and the Python and R code used to clean, analyze, and model the data\footnote{\scriptsize https://gitlab.eecs.umich.edu/mdst/D4GX-2017-Detroit-Vehicles}. We hope that this will lead to further similar data explorations in other domains, extensions of our methodology. %and hopefully extends the methodology used here. 

\section{Future Work}
\label{sec:future}

There are several promising avenues for future research. While we apply the PARAFAC decomposition to $vehicle \times system \times time$ data tensors, nothing about this approach limits it to evaluating these three specific variables. This analysis should be applied to several other dimensions, including WAC and Job Descriptions, other measures of vehicle lifetime (i.e., mileage), garage (location) and technician. Future work can utilize this and other information to build more robust predictive models to predict demand,  maintenance costs, and vehicle downtime (repair duration), and to assess maintenance effectiveness. Furthermore, as is demonstrated by the wide variety of applications of tensor decompositions discussed in Section \ref{sec:related}, this approach could be extended to any civic data where complex multivariate relationships exist by using the open-source code provided.

% \enlargethispage{3\baselineskip}

\section*{Acknowledgements}

This work is partially supported by National Science Foundation, grant IIS-1453304. 
It would not have happened without the support of the broader Michigan Data Science Team. The authors recognize the support of Michigan Institute for Data Science (MIDAS) and computational support from NVIDIA. We would like to thank the General Services Department of the City of Detroit for bringing this project to our attention and making the data available for use.

{\small
\bibliographystyle{abbrv}
\balance
\bibliography{main}
}

\end{document}